# Design of Robust Controller Applied for Series Elastic Actuators in Controlling Humanoid's Joint

Anh Khoa Lanh Luu[1], Van Tu Duong[1,2], Huy Hung Nguyen[1,3], Sang Bong Kim[4] and Tan Tien Nguyen[1,*]

[1]*National Key Laboratory of Digital Control and System Engineering, Ho Chi Minh City University of Technology, VNU-HCM, Hochiminh city, Vietnam.*

[2]*Department of Mechatronics, Ho Chi Minh City University of Technology, VNU-HCM, Hochiminh city, Vietnam.*

[3]*Faculty of Electronics and Telecommunication, Saigon University, Vietnam*

[4]*Pukyong National University, Busan, Republic of Korea*

[*]*Corresponding author. E-mail: nttien@hcmut.edu.vn*

**Abstract**

Although the application of Series elastic actuators (SEAs) in the biomechatronic field has proved its appropriation in many aspects so far, the problems of maintaining the stability for the SEAs still remains. This paper proposes a robust controller so that to overcome the drawbacks of the previous researches. Firstly, a mathematical model considering both the SEAs and the hip joint of humanoid UXA-90 is obtained. Secondly, a reference input of the proposed controller that is achieved from desired hip joint's angle in a walking cycle is utilized. Then, a backstepping based sliding mode force control approach is employed to ensure the precise movement of robot's link as well as meeting the requirement of robustness for the whole system, which is significant for the task of walking of a humanoid. Finally, some simulations are carried out to verify the quality and effectiveness of the proposed controller.

## 1    Introduction

Actuators are essential components in most modern robots since they provide the motive power for the robots. Electromagnetic, hydraulic and pneumatic actuators have been widely utilized in industry applications. With the advantages such as high precision, high torque and easy to control, electromagnetic actuators are used extensively in industrial robot that involve welding, painting, assembling and other repeatable tasks. Hydraulic actuators offer very high force/torque and speed characteristic which can outperform current electromagnetic actuators.

These types of actuators are commonly used in humanoid robots. DC Motors are used in the KHR series, HRP series, Humanoid Robot series from HONDA (Park, I. W. *et al.*, 2004) (Park, I. W. *et al.*, 2007) *(Kaneko, K. et al.*, 2002) (Akachi, K. *et al.*, 2005) (Kazuo, H *et al.*, 1988) (Sakagami, Y. *et al.*, 2002). Whereas Atlas from Boston Dynamic utilized hydraulic actuators for its lower limbs, torso and upper arms, while the forearms and wrists are electrically powered (Banerjee, N. *et al.*, 2015). TaeMu humanoid robot (Hyon, S. H. *et al.*, 2017) has fifteen active joints driven by hydraulic servo cylinders. Hydra humanoid robot (Ko, T. *et al.*, 2018) utilized Electro-Hydrostatic Actuator (EHA) to simultaneously realize both of the back drivability and high control bandwidth. Pneumatic actuator are often used in robot hands and artificial muscle (Kim, K. R. *et al.*, 2018) (Li, S. *et al.*, 2019) (Bierbaum, A *et al.*., 2009) (Nishino, S. *et al*., 2007) since they cannot generate large force or torque like electromagnetic and hydraulic actuators.

However, the advantageous stiffness property in these actuators could cause devastating effect on the robot's system themselves. Humanoid robots are expected to work in dangerous environment and cooperate with human in various condition working space, thus they are subjected to numerous external impact forces which could damage the motor through transmission elements. Various solutions to overcome this drawback have been proposed in the literature so far. One approach is to use compliance control using torque feedback. Another method is impedance control, it was introduced by Neville Hogan (Hogan, N., 1985) and was adopted in various researches for controlling rigid or flexible joints, *e.g* (Suarez, A *et al*., 2018) (Zhu, H *et al*., 2017) (Jiang, Z. H. *et al*., 2019,) (Hirayama, K. *et al*., 2019). Nevertheless, this control strategy is limited when controlling interaction force with the environment. Therefore, compliant actuator has received considerable attention.

The compliant actuator involves integrating an elastic element into a stiffness actuator. With proper design, this technique not only protects the entire system against impact, storing energy in the compliant element but also offers additional information about force/torque in joints of the robot. This design has been developed in various humanoid robot such as M2V2 (Pratt, J *et al*., 2012), ESCHER (Knabe, C *et al*., 2015), COMAN (Ajoudani, A *et al*., 2015), WALK-MAN (Tsagarakis, N. G. *et al.*, 2017).

In our previous studies, the commercial humanoid robot UXA-90 can only walk straight on a perfectly flat ground with a pre-calculated trajectory. The humanoid robot suffered significant impact force when walking, causing the vibration of the whole system. The impact force also causes great wear in the actuator's transmission gear over time. Therefore, it is crucial to implement the compliant actuator into UXA-90's legs since it will ease the impact force, eliminating vibration on the humanoid robot.

Since we have already succeeded in designing the SEA actuator for humanoid robot (Truong, K. D. *et al.*, 2020) (Truong, K. D. *et al.*, 2020).In this paper, SEA actuators are used for two legs instead of the original DC servo actuators. First, a dynamic model is deduced for one leg of the humanoid as well for the SEA. Secondly, a novel backstepping based sliding mode force control approach is employed to build the controller for each joint of the leg. Finally, some simulations are carried out to evaluate the effectiveness of the controller and the SEA mechanism.

## 2  System Modelling

### 2.1 Overall system modelling

The overall system is modelled as in Figure 1. The lower limb of humanoid is simplified to an L-shape link (E-D-C) which rotates about the hip joint (denoted as point E). The whole mass distribution of the limb is concentrated into one point (F) with the amplitude of $m$. The SEA has one end rotating about point B and the other end connecting to point C of the limb. Thus, the prismatic movement and linear force of the SEAs results in the rotation movement and external torque acting on the robot link.

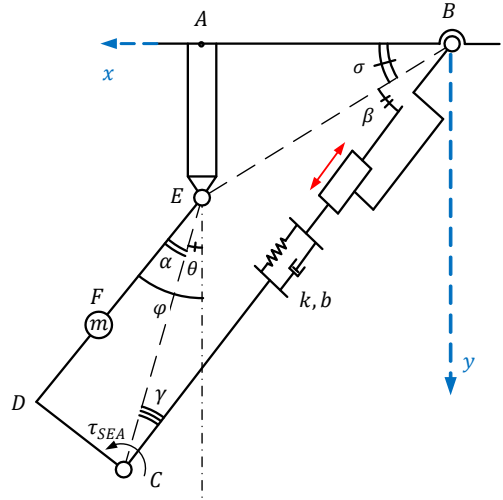

Figure 1.  Dynamic model of overall system.

The geometric parameters in the proposed dynamic model are denoted as below:
$$CD = d_1, DF = d_2, FE = d_3, EA = d_4, AB = d_5, CE = d_6, EB = d_7$$
The amplitude torque caused by gravity at point F is defined as:
$$\tau_R = mgd_3 \sin \varphi \tag{1}$$
This torque results in a reaction force acting along the SEA at point C, which is:
$$F_R = \frac{\tau_R}{d_6 \sin \gamma} \tag{2}$$
The length of SEA can be obtained by solving the geometric relationship of the SEA's system:
$$L(\theta) = \sqrt{d_4^2 + d_5^2 + d_6^2 + 2d_6(d_5 \sin \theta + d_4 \cos \theta)} \tag{3}$$
By defining the following:
$$\sin \gamma = \frac{d_7 \sin\left(\theta + \sigma + \frac{\pi}{2}\right)}{L(\theta)} \tag{4}$$

Eq. (2) can be rewritten as:
$$F_R = \frac{mgd_3 \sin(\theta + \alpha)}{d_6 d_7 \sin\left(\theta + \sigma + \frac{\pi}{2}\right)} L(\theta) \tag{5}$$

Define $\tau_{SEA}$ as the torque acting at point C by SEA's system. The relationship between $\varphi$ and $\theta$ is as followed:
$$\varphi = \alpha + \theta \tag{6}$$

where: $\alpha = arctan\left(\frac{d_1}{d_2+d_3}\right)$, $\theta$ is the angle between the $CE$ line and the line perpendicular to $x$ axis at point E, $\varphi$ is the joint's angle.

The dynamic of the system described in Figure 1 is given by:
$$md_3^2 \ddot{\varphi} = -mgd_3 \sin \varphi + \tau_{SEA} + \tau_D - B\dot{\varphi} \tag{7}$$

where $B$ is damping coefficient of the joint, $\tau_D$ is the system's disturbance torque, $\tau_{SEA}$ is the torque created by the SEA.

Since $B$ and $\tau_D$ are the system's uncertainties, in order to achieve the control objective, an integral sliding mode control (ISMC) is implemented to guarantee the robustness of the whole system despite the uncertainty parameters and external disturbances.

## 2.2 SEA modelling

For the mechanical aspect of the actuator, the dynamic equation of the output point $C$ is:
$$\ddot{X}_C + \frac{k}{m} X_C = \frac{k}{m} X_0 - \frac{F_R}{m} \tag{8}$$

in which, $k$ is the stiffness coefficient of the spring used in the SEAs, $X_C$ and $X_0$ are the displacements of the SEA's end effector and the ball screw's nut, respectively.

For the electrical aspect of the actuator, the characteristic of an ideal DC motor can be expressed by:
$$V_{IN} = IR + V_{EMF} + L\frac{dI}{dt} \tag{9}$$

where $V_{IN}$ is the DC motor's applied voltage, $I$ is the armature current, $R$ is the motor's internal resistance, $L$ is the inductance.

The back-EMF voltage defined as $V_{EMF} = K_{EMF}\omega_M$ ($K_{EMF}$ is the back-EMF constant) and the electrical torque defined as $T_M = K_T I$ ($K_T$ is the torque constant), Eq. (9) can be rewritten as:
$$V_{IN} = IR + K_{EMF}\omega_M + \frac{L}{K_T}\frac{dT_M}{dt} \tag{10}$$

By applying Newton - Euler equation for the motor's rotor, the dynamics of the rigid part can be written as:
$$J_M \dot{\omega}_M = T_M - \frac{1}{n\eta_1}\left(J_s \dot{\omega}_s + \frac{l}{2\pi\eta_2}(m_0 \dot{v}_0 + F_L)\right) - B_M \omega_M \tag{11}$$

where $J_M$, $\omega_M$ and $J_s$, $\omega_s$ are the moment of inertia and the angular velocity of the motor and ball screw respectively, $m_0$ and $v_0$ are the mass and velocity of the ball screw's nut together with the spring base, $B_M$ is the viscous friction coefficient, $n$ is the gearbox transmission ratio, $l$ is the lead of ball screw, $\eta_1$ and $\eta_2$ are the efficiency coefficients of the gearbox and the ball screw respectively, $F_L$ is the reaction force from the load.

For simplicity, define the equivalent moment of inertia of the whole SEA-system $J_{eq}$, which includes the moment of inertia of motor, ball screw, ball screw nut and spring base:
$$J_{eq} = J_M + \frac{1}{n^2 \eta_1} J_s + \frac{l^2}{4\pi^2 n^2 \eta_1 \eta_2} m_0 \tag{12}$$

Moreover, the torque acted on the load (which is also the reaction torque from the load) can be calculated from load force as:
$$T_L = \frac{l}{2\pi n \eta_1 \eta_2} F_L \tag{13}$$

Substituting Eq. (12) and Eq. (13) into Eq. (11), and recall that $\omega_M = v_0 \frac{2\pi n}{l}$, the mechanical equation of the rigid part can be represented as:
$$J_{eq} \frac{2\pi n}{l} \dot{v}_0 = T_M - T_L - B_M v_0 \frac{2\pi n}{l} \tag{14}$$

By substituting Eq. (14) into Eq. (10), the differential equation describes the system can be written as:

$$V_{IN} = \frac{T_M}{K_T}R + K_{EMF}\frac{n2\pi}{l}v_0 + \frac{L}{K_T}\frac{dT_M}{dt} \tag{15}$$

Expand the Eq. (15), it yields:

$$V_{IN} - \frac{RT_L - L\dot{T}_L}{K_T} = \frac{2\pi n}{lK_T}\left(LJ_{eq}\ddot{v}_0 + (RJ_{eq} + LB_M)\dot{v}_0 + (B_M R + K_{EMF}K_T)v_0\right) \tag{16}$$

$U_v^*$ is defined as:

$$U_v^* = V_{IN} - \frac{RT_L - L\dot{T}_L}{K_T} \tag{17}$$

Hence the dynamic equation of the SEAs is expressed as:

$$U_v^* = \frac{2\pi n}{lK_T}\left(LJ_{eq}\ddot{v}_0 + (RJ_{eq} + LB_M)\dot{v}_0 + (B_M R + K_{EMF}K_T)v_0\right) \tag{18}$$

The unknown parameters are determined by examining the step signal response in no-load state of the DC motor used in our project. The characteristic equation of just only the motor is:

$$V_{IN} = \frac{LJ_M\ddot{\omega}_M + (RJ_M + LJ_M)\dot{\omega}_M}{K_T} + \frac{(RB_M + K_{EMF}K_T)\omega_M}{K_T} \tag{19}$$

The System Identification toolbox and Simulink toolbox of MATLAB are utilized for this task. The results are: $R = 5,56\ \Omega$, $B_M = 16,5 \times 10^{-5}\ Nms/rad$, $J_M = 1,57 \times 10^{-4}\ Kgm^2$, $K_T = K_{EMF} = 0,202\ Nm/A$, $L = 4,6\ mH$.

The equivalent moment is then calculated as:

$$J_{eq} = J_M + \frac{1}{n^2\eta_1}J_G + \frac{l^2}{4\pi^2 n^2 \eta_1 \eta_2}m_0 = 1,574 \times 10^{-4}(Kgm^2) \tag{20}$$

Then, Eq. (18) becomes:

$$U_v^* = (4.69 \times 10^{-3})\ddot{v}_0 + 5.68\dot{v}_0 + 270v_0 \tag{21}$$

Since the coefficient of $\ddot{v}_0$ is relative small compared to others, it can be neglected to simplify the equation, which leads to:

$$U_v^* = 5,68\dot{v}_0 + 270v_0 \tag{22}$$

Let $U_v = U_v^*/5.68$, it yields:

$$U_v = \dot{v}_0 + 47,535 v_0 = \ddot{X}_0 + 47,535\dot{X}_0 \tag{23}$$

Combine Eq. (8) and Eq. (23), the dynamic equation of the SEA is obtained as:

$$(\ddot{X}_C - \ddot{X}_0) = -\frac{k}{m}(X_C - X_0) - \frac{F_R}{m} + 48\dot{X}_0 - U_v \tag{24}$$

Let $[\Delta] = X_C - X_0$, $\omega = \sqrt{\frac{k}{m}}$, $U_{eq} = 48\dot{X}_0 - U_v$, Eq. (24) becomes:

$$[\ddot{\Delta}] + \omega^2[\Delta] = U_{eq} - \frac{F_R}{m} \tag{25}$$

## 3    Controller Design

### 3.1 Sliding mode approach

First, the torque created by the SEA is calculated as:

$$\tau_{SEA} = F_{SEA}\frac{d_6 d_7 \sin\left(\varphi + \sigma + \frac{\pi}{2} - \alpha\right)}{L(\varphi)} = -[\Delta]\frac{kd_6 d_7 \sin\left(\varphi + \sigma + \frac{\pi}{2} - \alpha\right)}{L(\varphi)} \tag{26}$$

where $F_{SEA} = -k[\Delta]$ is the output force of the SEA, $k$ is the total stiffness coefficient of the springs inside the SEA Then, Eq. (7) is rewritten as:

$$\ddot{\varphi} = \frac{-mgd_3 \sin\varphi - B\dot{\varphi} + \tau_D}{md_3^2} - [\Delta]\frac{kd_6 d_7 \sin\left(x_1 + \sigma + \frac{\pi}{2} - \alpha\right)}{md_3^2 L(x_1)} \tag{27}$$

From Eq. (27), the system state variables can be defined as follow:

$$\begin{aligned} x_1 &= \varphi \\ \dot{x}_1 &= x_2 \\ \dot{x}_2 &= u.g(x_1) + f(x_1, x_2, \tau_D) \end{aligned} \tag{28}$$

in which: $u \triangleq [\Delta]$, $f(x_1, x_2, \tau_D) \triangleq -\frac{Bx_2 + mgd_3 \sin x_1 - \tau_D}{md_3^2}$, $g(x_1) \triangleq -\frac{kd_6 d_7 \sin\left(x_1 + \sigma + \frac{\pi}{2} - \alpha\right)}{md_3^2 L(x_1)}$

The first error variable is defined as the differences between the desired and response rotation angle while the second one is between the desired and response angular velocity. They are shown as:

$$e_1 = \varphi_d - \varphi \qquad (29)$$
$$e_2 = \dot{e}_1$$

The sliding function is determined as:
$$\sigma = e_2 + ce_1 \quad , c > 0 \qquad (30)$$

The first time-derivative of sliding function yields:
$$\dot{\sigma} = \dot{e}_2 + c\dot{e}_1 = \ddot{\varphi}_d - g(x_1)u - f(x_1, x_2, \tau_D) + c(\dot{\varphi}_d - \dot{\varphi}) \qquad (31)$$

The Candidate Lyapunov Function is chosen as:
$$V_x = \frac{1}{2}\sigma^2 \qquad (32)$$

Take the first time-derivative of CLF, it yields:
$$\dot{V}_x = \sigma\dot{\sigma} = \sigma[\ddot{\varphi}_d - u.g(x_1) - f(x_1, x_2, \tau_D) + c(\dot{\varphi}_d - \dot{\varphi})] \qquad (33)$$

To ensure the stabilization of the system, the following condition must be met $\dot{V}_x \leq 0$. For that requirement, the control signal $u$ must be determined as:
$$u = \frac{1}{g(x_1)}[\rho \, sign(\sigma) + \ddot{\varphi}_d - f(x_1, x_2, \tau_D) + c(\dot{\varphi}_d - \dot{\varphi})] \qquad (34)$$

where $\rho$ is a positive constant. The time-derivative of CLF becomes:
$$\dot{V}_x = -\rho \, \sigma \, sign(\sigma) = -\rho|\sigma| \leq 0 \qquad (35)$$

This leads to the fact that $\dot{V}_x$ and $\sigma$ are bounded. Moreover, $\dot{V}_x$ is an absolute function so that it is uniform continuity. Hence, $\dot{V}_x \to 0$ when $t \to \infty$ and $\sigma$ is proven to converge to zero.

### 3.2 Backstepping algorithm

In this section, the dynamic system in Eq. (27) is combined with Eq. (25) to determine the function of input voltage $U_{eq}$ which is the primary control variable.

First, Eq. (27) is express in the state space form as follow:
$$\dot{X} = f_x(X) + g_x(X)u \qquad (36)$$

where $X \triangleq [x_1 \quad x_2]$, $f_x(X) \triangleq \begin{bmatrix} x_2 \\ -\frac{Bx_2 + mgd_3 \sin x_1 - \tau_D}{md_3^2} \end{bmatrix}$, $g_x(X) \triangleq \begin{bmatrix} 0 \\ -\frac{kd_6 d_7 \sin\left(x_1 + \sigma + \frac{\pi}{2} - \alpha\right)}{md_3^2 L(x_1)} \end{bmatrix}$

Consider the first augmented dynamic system:
$$\begin{cases} \dot{X} = f_x(X) + g_x(X)z_1 \\ \dot{z}_1 = u_1 \end{cases} \qquad (37)$$

In the previous section, an adaptive control law $u_x = u$ is proven to be able to ensure the stability of the X-system. The control goal now is to make $z_1$ follow the desired value of $u_x$. The augmented Lyapunov function is:
$$V_1 = V_x + \frac{1}{2}(u_x - z_1)^2 = \frac{1}{2}\sigma^2 + \frac{1}{2}(u_x - z_1)^2 \qquad (38)$$

The first time-derivative of $V_1$ yields:
$$\dot{V}_1 = \sigma[\ddot{\varphi}_d - z_1 g(x_1) - f(x_1, x_2, \tau_D) + c(\dot{\varphi}_d - \dot{\varphi})] + (u_x - z_1)(\dot{u}_x - u_1) \qquad (39)$$

By adding and subtracting the term $\sigma g(x_1)u_x$, the Eq. (39) becomes:
$$\dot{V}_1 = \sigma[\ddot{\varphi}_d - u_x g(x_1) - f(x_1, x_2, \tau_D) + c(\dot{\varphi}_d - \dot{\varphi})] + \sigma g(x_1)(u_x - z_1) + (u_x - z_1)(\dot{u}_x - u_1) \qquad (40)$$

To stabilize the proposed extended system, the pseudo-control law is:
$$u_1 = k_1(u_x - z_1) + \dot{u}_x + \sigma g(x_1) \qquad (k_1 > 0) \qquad (41)$$

Continue the back-stepping algorithm, the second augmented system is proposed as:
$$\begin{cases} \dot{X}_1 = f_1(X_1) + g_1(X_1)z_2 \\ \dot{z}_2 = f_2(X_2) + g_2(X_2)U_{eq} \end{cases} \qquad (42)$$

in which: $X_1 \triangleq \begin{bmatrix} X \\ z_1 \end{bmatrix}$, $f_1(X_1) \triangleq \begin{bmatrix} f_x(X) + g_x(X)z_1 \\ 0 \end{bmatrix}$, $g_1(X_1) \triangleq \begin{bmatrix} 0 \\ 1 \end{bmatrix}$, $X_2 \triangleq \begin{bmatrix} X_1 \\ z_2 \end{bmatrix}$, $f_2(X_2) \triangleq -\omega^2 z_1 - \frac{F_R}{m}$, $g_2(X_2) \triangleq 1$

The $X_1$-system is assumed to be stable by the control law $u_1$. The goal is to make $z_2$ reach the value of $u_1$. The $2^{nd}$ augmented Lyapunov function:

$$V_2 = V_1 + \frac{1}{2}(u_1 - z_2)^2 \tag{43}$$

Take the first time-derivative of Eq. (43), it yields:

$$\dot{V}_2 = \sigma[\ddot{\varphi}_d - u_x g(x_1) + f(x_1, x_2, \tau_D) + c(\dot{\varphi}_d - \dot{\varphi})] + \sigma g(x_1)(u_x - z_1) + (u_x - z_1)(\dot{u}_x - z_2) \\ + (u_1 - z_2)[\dot{u}_1 - f_2(X_2) - g_2(X_2)U_{eq}] \tag{44}$$

By adding and subtracting the term $(u_x - z_1)u_1$, Eq. (44) becomes:

$$\dot{V}_2 = \sigma[\ddot{\varphi}_d - u_x g(x_1) + f(x_1, x_2, \tau_D) + c(\dot{\varphi}_d - \dot{\varphi})] + \sigma g(x_1)(u_x - z_1) + (u_x - z_1)(\dot{u}_x - u_1) \\ + (u_x - z_1)(u_1 - z_2) + (u_1 - z_2)[\dot{u}_1 - f_2(X_2) - g_2(X_2)U_{eq}] \tag{45}$$

Then, control input for the second augmented system, which is the input voltage, is obtained as:

$$U_{eq} = \frac{1}{g_2(X_2)}[-f_2(X_2) + k_2(u_1 - z_2) + \dot{u}_1 + (u_x - z_1)] \qquad (k_2 > 0) \tag{46}$$

## 4  Simulation and Analyzing

The simulation is carried out to evaluate the capability of the control algorithm. The physical and geometric parameters of the system are given in Table I.

**Table I. Simulation parameters**

| Physical parameters | |
|---|---|
| Mass, $m$ (kg) | 2 |
| Viscous friction coefficient, $B$ ($Nms$) | 0.5 |
| Stiffness coefficient, $k$ ($N/m$) | 20000 |
| Geometric parameters | |
| $d_1$ ($m$) | 0.0280 |
| $d_2$ ($m$) | 0.0525 |
| $d_3$ ($m$) | 0.0525 |
| $d_4$ ($m$) | 0.0350 |
| $d_5$ ($m$) | 0.1180 |

The referenced signal used is the rotation angle at the hip joint of humanoid UXA-90 in a walking cycle, which is inferred from (Nguyen, X. T. *et al.*, 2020).

First, the sliding mode and back-stepping controller are designed with the initial gain values chosen as: $c = 10$, $\rho = 3$, $k_1 = 1$, $k_2 = 5$. The results are shown in Figure 2.

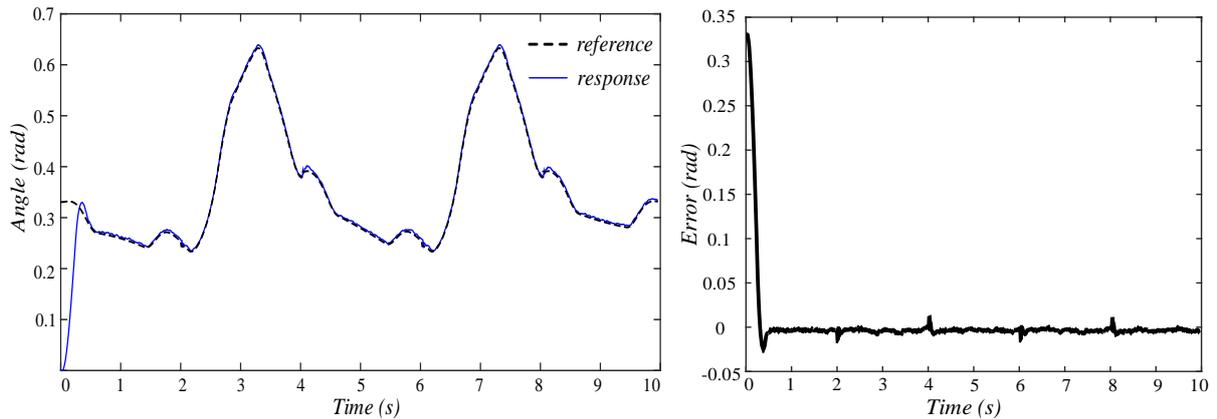

**Figure 2.** The simulated response in comparison with the referenced signal (left) and the angle tracking error (right)

Overall, the output response of the system can track the referenced angle trajectory with relative small error. In specific, except for the initial error during the transient phase, the highest absolute error is approximate $0.02\ rad$, which is less than 10% compared to the tracking value in every instants during the whole cycle. The disturbances mainly appear at the peak of the graph since it is the moment when the joint reverses its rotating direction rapidly.

Next, the influence of each gain coefficients in the sliding mode based backstepping control law is examined. Figure 3 illustrates the simulated responses of the system in different sets of gain values. In each cases, each coefficients is modified respectively while the others are kept unchanged. Based on the obtained results in Figure 3, the impact of $c, \rho, k_1, k_2$ can be summarized. First, the increases of $k_1, k_2$ and $\rho$ reduce the transient time and changing-phase instant of the system while the increase of $c$ causes a contrast effect. However, the higher the value of $c$, the lower the steady-state error. Besides, high values of $k_1$ and $k_2$ lead to the overshoot of the response, which can make the system become unstable. The increase of $\rho$ hardly create overshoot phenomenon like $k_1$ and $k_2$. However, a control law with high value of $\rho$ will make the system stay at high steady-state error.

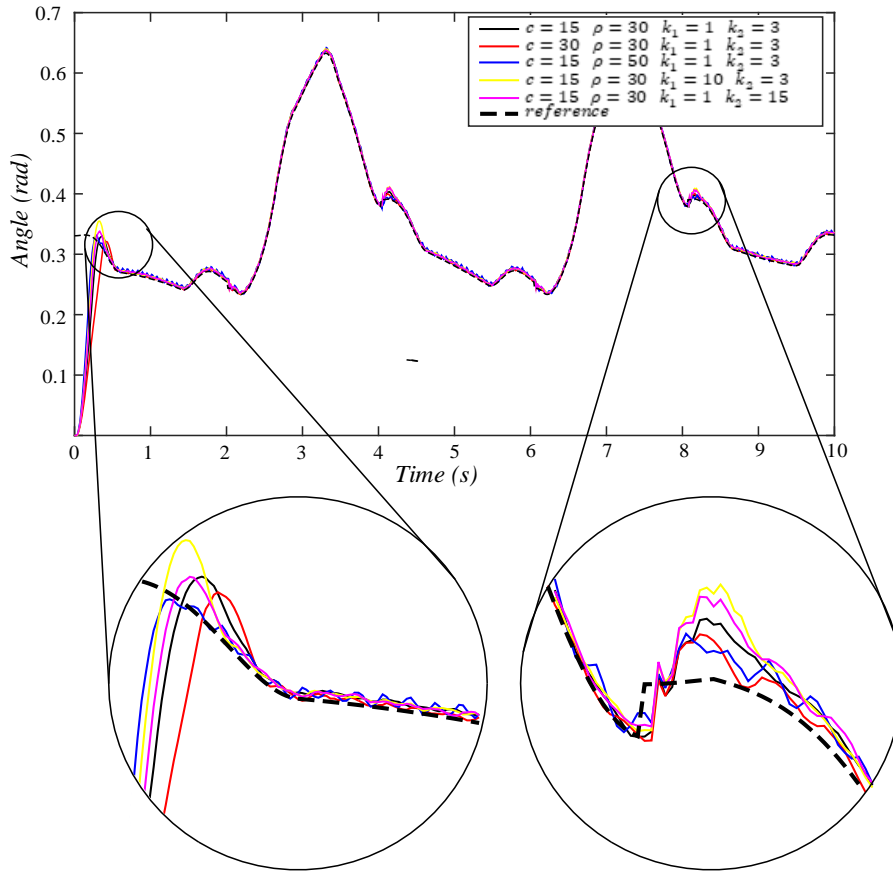

**Figure 3. The simulated response in different sets of gain value in the control law.**

The reasons for those impacts mainly rely on the positions and roles of each coefficients in the proposed equations. Coefficient $c$ is defined in the sliding function in Eq. (30) and it can be considered as a pole of the error system $(e_1, e_2)$, which explains why the increase of $c$ leads to the stabilization with low overshoot of the output variable $\varphi$ of the system. Meanwhile, $\rho$ appears as the coefficient of term $sign(\sigma)$ in the control law $u$. Since the output values of this term only vary among definite numbers (-1; 0 and 1), $\rho$ has no impact to the variation of $\sigma$ except when it changes sign, which means $\rho$ just assures the bounded error for the system with the same recovery speed (equal to $\rho$ itself). However, due to the constant recovery speed at every error value, the high value of $\rho$ will make the system's response oscillates in the steady-state about the zero-error point and with a high enough value, it would lead to the unstable response. In the other hand, $k_1$ and $k_2$ are the proportion coefficients which influence the errors of $z_1$ and $z_2$ respectively. For that reason, they perform just like the $K_p$ coefficient in a standard PID controller. In specific, the higher values of them will make the system rapidly recover whenever the error occurs, however it may also cause huge overshoot and make the system unstable if assigned by too high values.

## 5   Conclusion

In conclusion, the sliding mode based back-stepping control algorithm is proven to be appropriate for this application. The effectiveness and quality of the controller is relied on the selections of gain values which can cause both positive and negative impact to the stability of the system. Moreover, although the simulation shown good results in this process, the real response of the system in practice might be different due to the incorrect measurements of the uncertainties parameters. In the future, some experiment will be carried out to further evaluate the capability of this control algorithm to the humanoid field.

## Acknowledgments

This research is supported by DCSELab and funded by Vietnam National University HoChiMinh City (VNU-HCM) under grant number B2019-20-09 and TX2021-20b-01.